# User Effects in Beam-Space MIMO

Mohsen Yousefbeiki, Halima Najibi, and Julien Perruisseau-Carrier

*Abstract*—The performance and design of the novel single-RF-chain beam-space MIMO antenna concept is evaluated for the first time in the presence of the user. First, the variations of different performance parameters are evaluated when placing a beam-space MIMO antenna in close proximity to the user body in several typical operating scenarios. In addition to the typical degradation of conventional antennas in terms of radiation efficiency and impedance matching, it is observed that the user body corrupts the power balance and the orthogonality of the beam-space MIMO basis. However, capacity analyses show that throughput reduction mainly stems from the absorption in user body tissues rather than from the power imbalance and the correlation of the basis. These results confirm that the beam-space MIMO concept, so far only demonstrated in the absence of external perturbation, still performs very well in typical human body interaction scenarios.

## I. INTRODUCTION

A MAIN impediment to the implementation of classical multiple-input multiple-output (MIMO) has been the difficulty of dealing with multiple RF chains and multiple antennas in low-end user terminals with complexity and size constraints. To tackle this challenge, there has recently been a considerable research effort on reduced-complexity and antenna-decoupling schemes to preserve the MIMO benefits while lowering the number of RF chains and/or decreasing physical dimensions. Among all existing techniques, the so-called *beam-space* MIMO has attracted a great deal of interest in wireless communications as the only solution in data multiplexing using a single RF chain and a compact reconfigurable radiating structure [1]-[8]. In contrast with traditional multiplexing transmission where each data sub-stream is independently assigned to a distinct RF chain/antenna connection, the beam-space multiplexing concept is based on mapping the MIMO information onto an orthogonal set of virtual basis patterns $B_n(\theta,\varphi)$ in the beam-space domain of a single antenna [1]. This is achieved by engineering the antenna instantaneous radiated field $E_{inst}(\theta,\varphi,t)$ such that at any instant $E_{inst}(\theta,\varphi,t) = \sum s_n(t) B_n(\theta,\varphi)$, where $s_n(t)$ are arbitrary complex data sub-streams to be multiplexed over the air. However, translating the beam-space MIMO concept to reality requires dealing with various challenges that are commonly overlooked in purely conceptual designs. Recent works have already addressed some of these challenging aspects such as the compatibility with portable devices [5]-[6], the emulation of higher-order modulation schemes [7], and the adaptation to non-ideal channel propagation conditions [8].

Another important challenge of practical beam-space MIMO relates to the inevitable variations of the antenna characteristics caused by the interaction of the radiated fields with surrounding objects and in particular the user. In the case of conventional user terminals, comprehensive studies have been conducted to investigate the user hand and head effects [9]-[13]. In beam-space MIMO, some of the MIMO sub-streams are directly modulated onto the antenna far-field. Therefore, the user's presence not only causes usual effects on the radiation efficiency and the impedance matching, but also affects the spatial multiplexing performance of the beam-space MIMO antenna. This makes the study of the user effect on beam-space MIMO distinct from those carried out in the case of conventional MIMO, and this issue is addressed here for the first time.

The remainder of this letter is organized as follows. In Section II, the beam-space MIMO concept is briefly recalled. Then, Section III discusses the effects of the user's presence on beam-space MIMO operation. In Section IV, the user effects are evaluated for different operating scenarios. Then, the role of different antenna parameters on channel capacity reduction is discussed. Finally, the impact of tuning the antenna reactive loads to take into account the user effect at the design level is examined.

## II. RECALL ON BEAM-SPACE MIMO

This study is based on the antenna shown in Fig. 1(a). This is a compact built-in antenna system designed in [6] for optimum single-radio multiplexing of BPSK modulated signals in *free space*. In order to facilitate the understanding of upcoming results, here we briefly describe the antenna operation and introduce the key performance parameters of beam-space MIMO.

In general, the antenna under study can be seen as a special realization of the antenna system symbolically represented in Fig. 1(b). It comprises a symmetric three-port radiator and two reactive loads $jX_1$ and $jX_2$. The reactive loads have been realized using two p-i-n diodes with opposite bias in the antenna under study. The central port (port 0) is the active input and matched to a 50-Ω excitation, while the other two are passive and terminated with two reactive loads. As the antenna system state is changed by swapping the reactive loads $[X_1, X_2] \rightarrow [X_2, X_1]$ (equivalent to altering the states of both p-i-n diodes in the antenna under study), the symmetry ensures constant impedance matching at the active port, and the instantaneous radiated field of the antenna system is mirrored with regard to the plane of symmetry, i.e. $E_I(\theta,\varphi) \rightarrow E_{II}(\theta,\varphi) = E_I(\theta,-\varphi)$ for a plane of

This work was supported by the Swiss National Science Foundation (SNSF) under grant n°133583.

The authors are with the group for Adaptive MicroNanoWave Systems, Ecole Polytechnique Fédérale de Lausanne (EPFL), CH-1015 Lausanne, Switzerland (e-mails: mohsen.yousefbeiki@epfl.ch, julien.perruisseau-carrier@epfl.ch).

symmetry at $\varphi=0-\pi$. Having two mirrored instantaneous radiation patterns guarantees the orthogonality of the angular patterns defined as

$$B_{1,2}(\theta,\varphi) := \left[E_{II}(\theta,\varphi) \pm E_{I}(\theta,\varphi)\right]/\sqrt{2} \qquad (1)$$

ensuring in turn the independency and the identical distribution of MIMO sub-channels in rich scattering environments [4]. Therefore, when the active port is fed by the first data stream $s_1$, it is possible to simultaneously transmit another BPSK signal $s_2$ by swapping the loads such that (see [4])

$$\begin{aligned} E_{inst}(\theta,\varphi,t) &= s_1(t) E_{I/II}(\theta,\varphi) \\ &= \left[s_1(t) B_1(\theta,\varphi) + s_2(t) B_2(\theta,\varphi)\right]/\sqrt{2}. \end{aligned} \qquad (2)$$

In addition to the basis orthogonality, an equal power distribution between the sub-streams is desired for open-loop MIMO operation. In the case of beam-space MIMO, this is equivalent to a power imbalance ratio of unity between the basis patterns $B_1$ and $B_2$. Moreover, an acceptable level of the total efficiency (product of radiation and matching efficiencies) at the active port is required. Optimum MIMO operation of the antenna system in terms of the power imbalance ratio and/or the total efficiency can be achieved by optimizing either the reactive loads [4] or the three-port radiator [6]. In summary, the performance of a beam-space MIMO antenna can be examined in terms of four key characteristics:

- Impedance matching,
- Radiation efficiency,
- Power balance of the basis,
- Correlation of the basis.

### III. USER EFFECTS IN BEAM-SPACE MIMO

The user effects on beam-space MIMO operation can be classified along two main categories. First, beam-space MIMO suffers from the *typical* effects encountered in any SISO or MIMO antenna system, namely, the variations of the impedance matching and the radiation properties, which are quite well-studied [9]-[13]. More specifically, a considerable portion of the power radiated is dissipated in the user tissues, which results in degradation of the radiation efficiency. The matching efficiency is also altered due to detuning, but this effect is generally less pronounced than the absorption losses.

The second category consists of the user effects on the basis correlation and the basis power imbalance ratio, which are *specific* to beam-space MIMO. It is obvious from (1) that the time-varying changes in the far-fields $E_I(\theta,\varphi)$ and $E_{II}(\theta,\varphi)$ lead to variations of the basis patterns $B_1$ and $B_2$ and the power imbalance ratio between them. On the other hand, the symmetry of the radiating environment no longer exists due to near-field coupling between the antenna and the user. This means that the antenna is no more capable of creating mirrored instantaneous radiation patterns. Therefore, the MIMO sub-channels become correlated since the orthogonality of the basis patterns $B_1$ and $B_2$ is no longer guaranteed. Furthermore, the impedance matching and the radiation efficiency become different between the two antenna states due to the introduced asymmetry.

Finally, it is worth noting that secondary user effects are not particular to beam-space MIMO. Conventional MIMO also suffers from its own specific effects such as variations of the power balance and the isolation among multiple antennas in human body interaction scenarios.

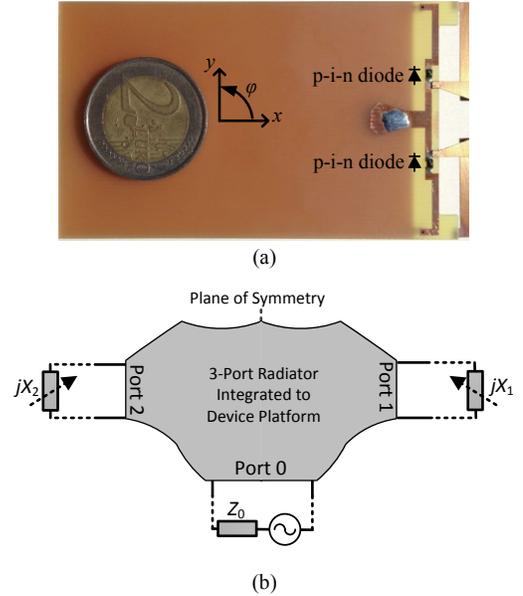

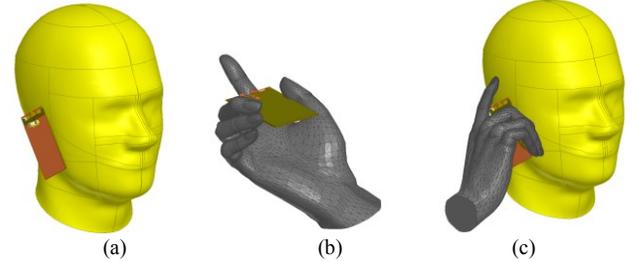

Fig. 1. (a) Beam-space MIMO antenna used in this study, designed and fabricated in [6]. (b) Symbolic modeling of the generalized antenna.

Fig. 2. The placement of the beam-space MIMO antenna under study in different scenarios: (a) nearby SAM head (cheek position), (b) in SAM hand (browsing mode), and (c) in SAM hand nearby SAM head (talking mode).

### IV. RESULTS AND DISCUSSION

#### A. User Body Modeling

The efficiency and the reliability of commercial EM full-wave simulations for evaluating user body effects have been demonstrated in several recent studies (see for instance [10]-[12]). In this work, full-sized standard hand and head phantoms, referred to as the specific anthropomorphic mannequins (SAM), are used for modeling the human body. The average dielectric properties of all corresponding tissues are selected according to [14]. The user effects are evaluated in the three operating scenarios shown in Fig. 2. The full-wave simulations have been performed using Ansys HFSS.

#### B. User Effect Evaluation

Fig. 3 shows the variations of the antenna reflection coefficient for different cases of the user body interaction. As expected, the impedance matching of the antenna slightly changes when placed near the body. Moreover, as mentioned in Section III, the reflection coefficient at the active port is different between the two system states due to the asymmetry of the radiating environment. However, these variations are rather insignificant as the return loss remains better than 10 dB over the operation frequency range.

The variations of the radiation properties are summarized in Table I. As with traditional antennas, the total efficiency of the antenna is severely degraded when placed in close proximity to the user body, dropping down from 77% in free-space case to 9% in talking mode. Since according to Fig. 3 the mismatch loss is almost unchanged, the efficiency reduction can be attributed to the absorption in body tissues.

It is observed that the parameters specific to beam-space MIMO are also affected. Power imbalance ratio between the basis patterns is affected by the presence of the user and varies from 1.1 to 4.1 for different case studied. Moreover, it is seen from Table I that the user body affects the orthogonality of the basis patterns. The basis correlation is increased from 0 ($-\infty$ dB) in free space to 0.3 (-10.5 dB) in talking mode, which is still acceptable for MIMO operation. Table I also shows the effect of the distance $d$ between the antenna and the head, where as expected smaller distances result in higher degradations.

Furthermore, capacity analyses were carried out in the different scenarios. We assumed the transmission of two BPSK signals using the beam-space MIMO antenna over a Kronecker narrowband flat-fading channel and the reception using two classic MIMO receivers. Fig. 4 shows the channel capacity of the MIMO system with respect to the transmit signal to noise ratio (SNR). It is observed that the system throughput is severely degraded in the presence of the user particularly in the low-SNR regime. The channel capacity is decreased by 41% from 1.7 b/s/Hz in free space to 1 b/s/Hz in the talking scenario for a SNR of 10 dB. Obviously, for high SNR the effect of the impairments is less pronounced, thus the curves approach each other.

It is also informative to evaluate the achievable multiplexing gain in the different operating scenarios. The multiplexing gain here measures the ratio between the channel capacity of the MIMO system and that of a SISO system with the same average total efficiency. As shown in Fig. 5, the multiplexing capability of the MIMO system is not much affected by the user's presence in most cases. In the worst case (near head, $d$ = 5 mm), where the specific parameters of beam-space MIMO are severely affected, the gain reduction is less than 23%. However, in general the variation is only about 5%-10%.

Finally, it is important to assess the contribution of each parameter (impairment) to the total performance degradation. To this aim, we selected the near head case (see Fig. 2a) as the reference, and compared the multiplexing efficiency [16] of the MIMO system in three new cases. In each case, only one of the antenna parameters (i.e. average total efficiency, power imbalance ratio, and basis correlation) was changed while the other two were left constant. The results are shown in Fig. 6, where it is immediately observed that the main contributor is the antenna efficiency. While the multiplexing efficiency is significantly affected by doubling the total efficiency, the contribution of halving the basis power imbalance ratio is less pronounced especially in lower SNR. The graph also shows that a little multiplexing efficiency enhancement is achieved when the correlation of non-ideal basis patterns is decreased by 3 dB.

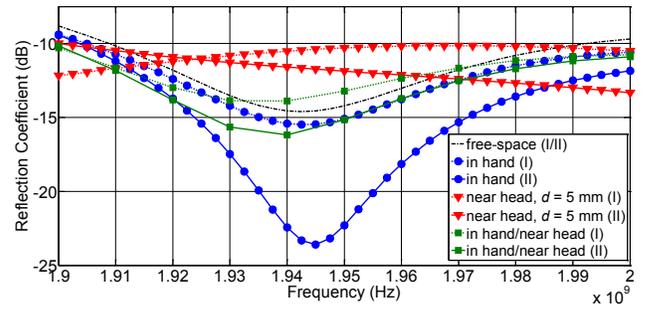

Fig. 3. The reflection coefficient of the beam-space MIMO antenna under study. The number inside the parentheses indicates the state of the antenna system. The parameter $d$ determines the minimum spacing between the antenna and the head.

TABLE I
USER BODY EFFECT ON BEAM-SPACE MIMO ANTENNA PERFORMANCE

| Case Study | Average Total Efficiency $\eta$ | Power Imbalance Ratio $r$ | Basis Correlation $|\sigma|^2$ (dB) |
|---|---|---|---|
| free-space | 77% | 1.3 | $-\infty$ |
| near head, $d$ = 5 mm | 13% | 4.1 | $-6.8$ |
| near head, $d$ = 7.5 mm | 18% | 2.0 | $-12.1$ |
| near head, $d$ = 10 mm | 24% | 1.2 | $-16.4$ |
| in hand | 40% | 1.6 | $-19.2$ |
| in hand/near head, $d$ = 10 mm | 9% | 1.1 | $-10.5$ |

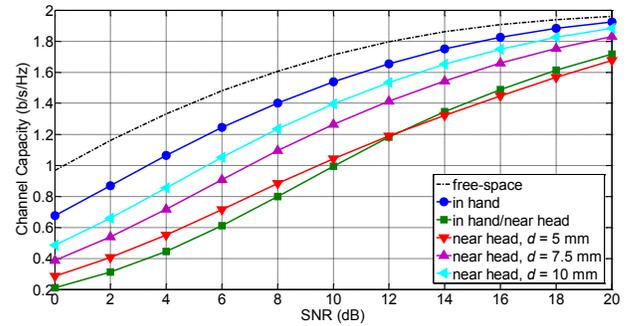

Fig. 4. BPSK capacity versus the transmit SNR in different scenarios. The simulation was performed based on Monte-Carlo method (see [15]) for an input of 10000 uniformly distributed BPSK symbols where 10000 channel realizations were used.

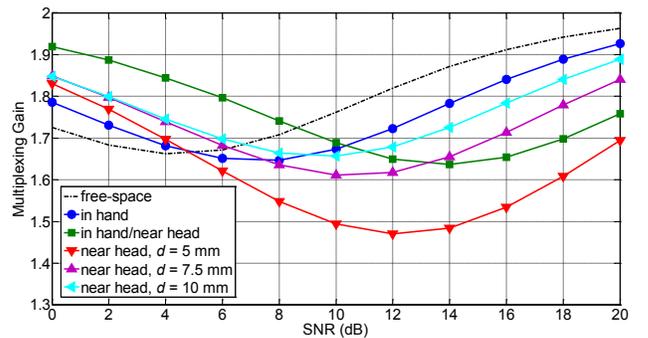

Fig. 5. Multiplexing gain versus the transmit SNR in different scenarios.

These results demonstrate that throughput reduction is mainly caused by the absorption in user body tissues rather than by the impairment of the basis power imbalance ratio and the basis correlation. Therefore, we can conclude that beam-space MIMO is not more vulnerable to the user effects than conventional MIMO systems, and has the potential to perform very well in typical human body interaction scenarios.

## C. Adaptive reactive loads

A possible way of decreasing the user effects is to adapt the antenna parameters in real time to the immediate environment, or in a lower complexity to take into account the user's presence in an average way a priori in the antenna design procedure. Since the beam-space MIMO antenna naturally embeds variable reactive loads, this can be realized by selecting the loads in such a way that maximum possible channel capacity in each operating scenario is achieved.

To this aim, we computed the channel capacity for a 2D range of reactive loads $X_1$ and $X_2$ under assumptions identical to those in Section IV.B. Fig. 7 shows a contour plot of the channel capacity for the browsing scenario (see Fig. 2b) where the reactance values of the original and adaptive designs are indicated by '+' and '×', respectively. Although the optimal loads are different from the reactance values of the p-i-n diodes, the improvement achieved is negligible. Table II summarizes changes of the channel capacity for each operating scenario. The use of adaptive reactive loads improves the channel capacity by only less than 3% while extra computational cost and hardware complexity is added to the beam-space MIMO system. This is fully in agreement with the results obtained in Section IV.B since adapting reactive loads mainly affects the basis power imbalance ratio and the basis correlation rather than the total efficiency.

## V. CONCLUSION

A first study of beam-space MIMO operation in the presence of the user has been presented. In addition to usual user effect such as total efficiency reduction, beam-space MIMO antennas suffer from specific user effects, namely, variations of the power imbalance ratio and the correlation between the basis patterns. However, such effects are insignificant as the performance degradation mainly stems from power dissipation in body tissues. Accordingly, adapting reactive loads of the beam-space MIMO antenna does not necessarily result in significant performance improvement. These results confirm that the beam-space MIMO concept, so far only demonstrated in the absence of external perturbation, still performs very well in typical human body interaction scenarios.

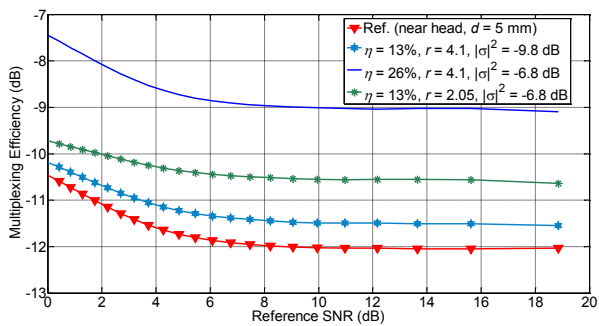

Fig. 6. Multiplexing efficiency variations versus antenna performance parameters.

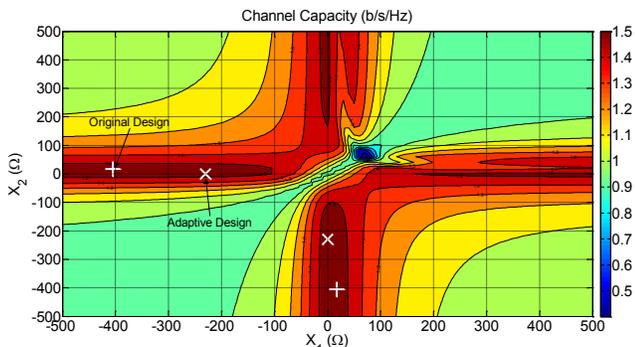

Fig. 7. Contour map of the channel capacity with respect to $X_1$ and $X_2$ for a SNR of 10 dB in browsing mode.

TABLE II
LOAD OPTIMIZATION EFFECT ON CHANNEL CAPACITY, SNR = 10 dB

| Case Study | $C_0$ (b/s/Hz) | $C_{max}$ (b/s/Hz) | Relative Difference |
|---|---|---|---|
| near head, $d$ = 5 mm | 1.04 | 1.07 | 3% |
| near head, $d$ = 7.5 mm | 1.25 | 1.28 | 2.5% |
| near head, $d$ = 10 mm | 1.35 | 1.38 | 2% |
| in hand | 1.52 | 1.54 | 1.5% |
| in hand/near head, $d$ = 10 mm | 1.00 | 1.03 | 3% |

$C_0$ and $C_{max}$ are the channel capacity values when the antenna embeds the p-i-n diodes and the optimal loads, respectively.